\documentclass[conference]{IEEEtran}

\ifCLASSINFOpdf
\else
\fi

\usepackage{graphicx}
\usepackage{bm}
\usepackage{amsfonts}
\usepackage{amsmath}
\usepackage{amssymb}
\usepackage{times}
\usepackage{subfigure}
\usepackage{latexsym,bm,amsmath,amssymb} 
\usepackage{CJK}
\usepackage{hhline}

\usepackage{multirow} 
\usepackage{amsmath}
\usepackage{xcolor}
\usepackage{epstopdf}
\usepackage{cite}
\usepackage[noend]{algpseudocode}
\usepackage{algorithmicx,algorithm}
\usepackage[left=1.65cm,right=1.65cm,top=1.91cm,bottom=2.7cm]{geometry}
\begin{document}
%
\title{Analysis of Cross-Domain Message Passing for OTFS Transmissions}

\author{\IEEEauthorblockN{
Ruoxi Chong\IEEEauthorrefmark{1},
Shuangyang Li\IEEEauthorrefmark{2}, 
Zhiqiang Wei\IEEEauthorrefmark{3}, 
Michail Matthaiou\IEEEauthorrefmark{1}, \\
Derrick Wing Kwan Ng\IEEEauthorrefmark{4},
and
Giuseppe Caire\IEEEauthorrefmark{2}
}
\IEEEauthorblockA{
\IEEEauthorrefmark{1}Centre for Wireless Innovation (CWI), Queen's University Belfast, UK\\
\IEEEauthorrefmark{2}Faculty of Electrical Engineering
and Computer Science, Technical University of Berlin, Germany\\
\IEEEauthorrefmark{3}School of Mathematics and Statistics, Xi'an Jiaotong University, P. R. China\\
\IEEEauthorrefmark{4}School of Electrical Engineering and Telecommunications, University of New South Wales, Australia
}
\vspace{-10mm}
}

\maketitle

%
\IEEEpeerreviewmaketitle

\begin{abstract}
In this paper, we investigate the performance of the cross-domain iterative detection (CDID) framework with orthogonal time frequency space (OTFS) modulation, where two distinct CDID algorithms are presented. The proposed schemes estimate/detect the information symbols iteratively across the frequency domain and the delay-Doppler (DD) domain via passing either the \textit{a posteriori} or extrinsic information. 
Building upon this framework, we investigate the error performance by considering the bias evolution and state evolution. 
Furthermore, we discuss their error performance in convergence and the DD domain error state lower bounds in each iteration.
Specifically, we demonstrate that in convergence, the ultimate error performance of the CDID passing the \textit{a posteriori} information can be characterized by two potential convergence points. In contrast, the ultimate error performance of the CDID passing the extrinsic information has only one convergence point, which, interestingly, aligns with the matched filter bound. Our numerical results confirm our analytical findings and unveil the promising error performance achieved by the proposed designs.


\let\thefootnote\relax\footnotetext{This work was supported in part by the U.K. Engineering and Physical Sciences Research Council (EPSRC) (grant No. EP/X04047X/1). The work of M. Matthaiou was supported by a research grant from the European Research Council (ERC) under the European Union’s Horizon 2020 research and innovation programme (grant No. 101001331). }
\end{abstract}
\section{Introduction}

OTFS modulation has emerged as a disruptive solution for robust communication over high-mobility channels, as evidenced by the extensive literature, e.g.~\cite{Wei2021magzine,Li2021performance,Ruoxi2022achievable ,Ruoxi2022outage}. The success of OTFS lies in the DD domain symbol placement,  where the channel response is more sparse and robust~\cite{Wei2021magzine,li2022Part2,Hadani:WCNC:2017,LSY_THP}.
However, as the input-output relation of OTFS is characterized by various forms of convolution, depending on the underlying pulse shape~\cite{Shuangyang_DDCom}, OTFS generally relies on higher-complexity detection algorithms to achieve satisfactory error performance than orthogonal frequency division multiplexing (OFDM)~\cite{Li2021performance}. 
Therefore, the pursuit of low-complexity detection for OTFS has long been a research focus.
For instance, in~\cite{Raviteja2018interference}, a message-passing algorithm was developed, where the interference is treated as a Gaussian variable to reduce the complexity.
Additionally, a sum-product algorithm based on the Ungerboeck observation model was proposed for OTFS in~\cite{Dehkordi2023beamspace}. Specifically, this algorithm guarantees that the cycles in the factor graph have at least a girth of six, thereby enjoying a better error performance. 
Furthermore, a CDID framework for OTFS was proposed in~\cite{li2021cross}. The key feature of this framework is the cross-domain message passing via the corresponding unitary domain transformation. Although a preliminary study of CDID was provided in~\cite{li2021cross}, numerous critical design aspects, such as the iteration mechanism and convergence analysis, remain not fully explored and understood, requiring further investigation. 
The investigation of such details is of importance both practically and theoretically, since it could not only facilitate the system design but also enhance the understanding
of cross-domain message passing.

In this paper, we pursue a performance analysis for two types of CDID algorithms, considering the bias evolution and state evolution. Specifically, we consider two types of CDID algorithms that operate in the frequency domain and the DD domain iteratively, where a minimum mean square error (MMSE) estimator is adopted in the frequency domain and a symbol-by-symbol detector is applied in the DD domain. Particularly, the Type-I CDID algorithm passes the \textit{a posteriori} information across two domains, while the Type-II CDID algorithm passes the \textit{extrinsic} information. 
We show that both two algorithms can be approximately unbiased and the the error state lower bounds are derived based on the state evolution.
Furthermore, we study the error performance in convergence. Particularly, we show that the Type-I CDID has two possible convergence points. In contrast, the Type-II CDID exhibits only one possible convergence point, which aligns well with the matched filter bound. Our numerical results validate our analytical findings and demonstrate a promising error performance.

\emph{Notations:}
The superscripts $(\cdot)^{\rm{H}}$, $(\cdot)^{\rm{T}}$, and $(\cdot)^{-1}$ denote the Hermitian transpose, transpose, and inverse of a matrix, respectively; ${\rm{diag}} \{ \cdot \}$ returns the diagonal elements of a matrix; ${{{\bf{F}}_N}}$ denotes the normalized discrete Fourier transform (DFT) matrix of size $N\times N$; ${\bf I}_M$ represents the $M\times M$ identity matrix; ``$ \otimes $" denotes the Kronecker product operator; 
$|\cdot|$ returns the cardinality of a set;
$\mathbb{E}\{\cdot\}$ denotes the statistical expectation; ${\mathbb{C}}$ 
denotes the complex number field; 
$\simeq$ and $\propto$ denote approximately equal to and proportional equal to, respectively. 

\vspace{-1mm}
\section{System Model}
\vspace{-1mm}
We consider a Zak transform (ZT)-based OTFS implementation~\cite{Shuangyang_DDCom}. Let ${\bf X}$ of size $M\times N$ be the DD domain modulated symbol matrix for OTFS, where $M$ denotes the number of delay bins/sub-carriers and $N$ denotes the number of Doppler bins/time slots, respectively. 
By passing through the inverse discrete Zak transform (IDZT) module and inserting a $L_{\rm CP}$-length cyclic prefix (CP), the time domain OTFS symbol vector ${\bf{\tilde s}} \in {\mathbb{C}}^{(MN+L_{\rm CP})\times 1}$ is given by 
\vspace{-1mm}
\begin{equation}
{\bf{\tilde s}} 
= {\bf{A}}_{\rm{CP}} {\bf{s}}
= {\bf{A}}_{\rm{CP}}\left( {{\bf{F}}_N^{\rm{H}} \otimes {{\bf{I}}_M}} \right){\bf{x}},
\label{OTFS_transmit_withCP}
\vspace*{-0.5\baselineskip}
\end{equation}
where 
${{\bf{x}}}$ is the vectorized ${{{\bf{X}}}}$, and ${\bf{A}}_{{\rm{CP}}}$ is the CP addition matrix for OTFS. Specifically, we have  ${\bf{ A}}_{{\rm{CP}}} \buildrel \Delta \over = {\left[ {{\bf{G}}_{{\rm{CP}}},{{\bf{I}}_{MN}}} \right]^{\rm{T}}}$, where ${{\bf{G}}_{\rm{CP}}}$ of size $MN \times L_{\rm CP}$ includes the last $L_{\rm CP}$ columns of the identity matrix ${\bf{I}}_{MN}$~\cite{RezazadehReyhani2018analysis}. 

After applying the transmitter pulse shaping $p(t)$ and passing through a time-varying channel, we apply a matched filter with the receive shaping pulse ${p^{\star}}\left( t \right)$ at the receiver side. Then, the vector-form received signal ${\bf{\tilde r}}$ can be written as
\vspace{-1mm}
\begin{equation}
{\bf{\tilde r}} = \sum\nolimits_{i = 1}^P {{{\bf{G}}^{\left( i \right)}}} {{{\bf{\tilde s}}}}+{\bf w},
\label{OTFS_io_withCP_mtx}
\vspace*{-0.5\baselineskip}
\end{equation}
where $P$ is the number of independent resolvable paths and ${\bf w}$ is the additive white Gaussian noise (AWGN) process with the one-sided power spectrum density (PSD) $N_0$.
Moreover, ${{{\bf{G}}^{\left( i \right)}}}$ is the $i$-th path of the time domain effective channel, whose $\left( {m,n} \right)$-th element ${g_{m,n}^{\left( i \right)}}$ can be denoted by
\vspace{-2mm}
\begin{align}
g_{m,n}^{\left( i \right)} \buildrel \Delta \over = {h_i}{e^{j2\pi n{\nu _i}{T_s}}}A_p^{*}\left( {\left( {n - m} \right){T_s} + {\tau _i},{\nu _i}} \right).
\label{effective_channel_coef}
\vspace*{-0.7\baselineskip}
\end{align}
In~\eqref{effective_channel_coef}, $h_i \in {\mathbb{C}}$, $\tau _i$, and $\nu _i$ denote the path gain, delay shift, and Doppler shift corresponding to the $i$-th path, respectively.
Moreover, ${A_p}\left( {\tau ,\nu } \right)$ is the ambiguity function of ${p}\left( t \right)$ with respect to delay $\tau$ and Doppler $\nu$, defined as
\vspace{-2mm}
\begin{equation}
{A_p}\left( {\tau ,\nu } \right) \buildrel \Delta \over = \int_{ - \infty }^\infty  {p\left( t \right)} {p^{\star}}\left( {t - \tau } \right){e^{ - j2\pi \nu \left( {t - \tau } \right)}}{\rm{d}}t
\label{AF}.
\vspace*{-0.5\baselineskip}
\end{equation}

Let ${\bf{R}}_{{\rm{CP}}}$ be the CP reduction matrix for OTFS, which is of size $MN\times (L_{\rm CP}+MN)$, obtained by removing the first $L_{\rm CP}$ rows of ${\bf I}_{L_{\rm CP}+MN}$. 
As such, the received time domain OTFS symbol vector ${\bf{r}}$ after removing the CP is given by
\vspace{-2mm}
\begin{equation}
{{\bf{r}}} = {\bf{R}}_{{\rm{CP}}}\sum\nolimits_{i = 1}^{P} {{{\bf{G}}^{\left( i \right)}}} {\bf{A}}_{{\rm{CP}}}{{\bf{s}}} + {\bf{w}}.
\label{OTFS_io_withoutCP_mtx_time}
\vspace*{-0.5\baselineskip}
\end{equation}
Here, we slightly abuse our notations for simplicity and adopt $\bf w$ for the noise vector after CP removal.
Finally, with domain transformation using the DZT, we arrive at the DD domain input-output relation as
\vspace{-2mm}
\begin{equation}
{{\bf{y}}} \!=\!\! \sum\nolimits_{i = 1}^{P} {\left( {{{\bf{F}}_N} \otimes {{\bf{I}}_M}} \right){\bf{R}}_{{\rm{CP}}}{{\bf{G}}^{\left( i \right)}}} {\bf{A}}_{{\rm{CP}}}\left( {{\bf{F}}_N^{\rm{H}} \otimes {{\bf{I}}_M}} \right){{\bf{x}}}\! +\! {\bf{w}}.
\label{OTFS_io_withoutCP_mtx}
\end{equation}

According to~\eqref{OTFS_io_withoutCP_mtx}, we define the effective time domain channel and equivalent frequency domain channel  as
\begin{align}
&{\bf G} \triangleq   {\bf R}_{\rm CP}\sum\nolimits_{i = 1}^{P} {{{\bf{G}}^{\left( i \right)}}}{\bf A}_{\rm CP}, \quad {\rm and}
\label{OTFS_effective_channel_time}\\
&{\bf{H}} \triangleq  {{\bf{F}}_{MN}}{\bf{G}}{\bf{F}}_{MN}^{\rm{H}}.
\label{OTFS_effective_channel_freq}
\end{align}
Let ${\bf z}={\bf{F}}_{MN}{\bf s}$ be the equivalent frequency domain symbol vector. 
The received frequency domain symbol vector ${\bf q}$ is given by 
\vspace{-5mm}
\begin{align}
{\bf{q}} = {{\bf{F}}_{{MN}}}{\bf{r}} = {\bf{H}}{\bf{z}} + {\bf{w}}.
\label{OTFS_freq_mtx1}
\end{align}

\section{Cross-domain Iterative Detection for OTFS}
In this paper, we 
consider frequency domain estimation (FDE) and DD domain symbol-by-symbol detection based on the framework in~\cite{li2021cross}. Note that FDE is of practical interest when the channel exhibits a small Doppler shift.

\subsection{Type-I: CDID via Passing a posteriori Information}
\begin{figure}
\centering
\includegraphics[width=0.4\textwidth]{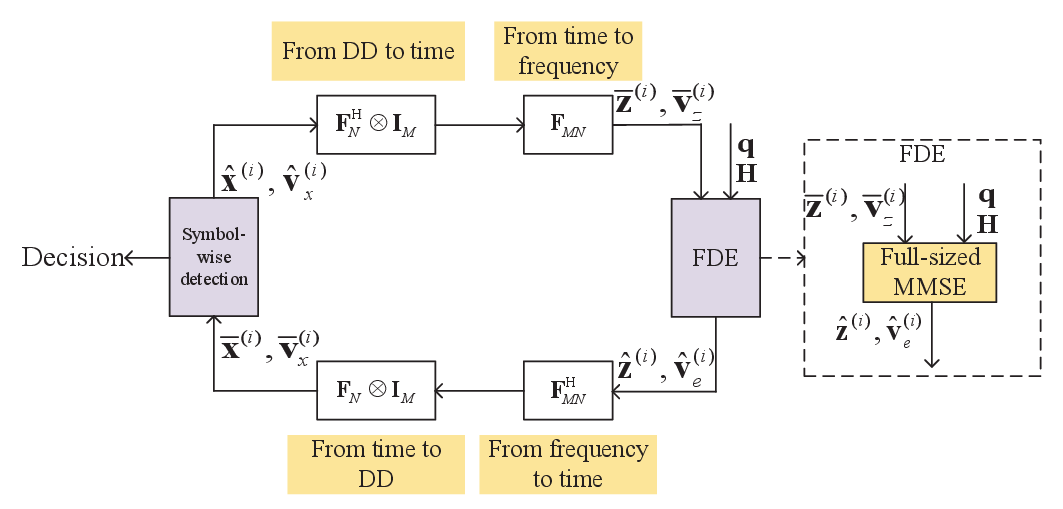}
\vspace{-3mm}
\caption{The block diagram for the Type-I CDID scheme.}
\label{cross_domain_typeI}
\centering
\vspace{-6mm}
\end{figure}
We first consider the CDID via passing \emph{a posteriori} information, as depicted in Fig.~\ref{cross_domain_typeI}. In the $i$-th iteration, the detection first starts from the frequency domain, whose inputs are 
${\bf q}$, 
${\bf{H}}$, the \emph{a priori} mean vector ${\bar {\bf z}}^{(i)}$, and variance vector ${\bar {\bf v}}_{z}^{(i)}$. Note that in the first iteration, the \emph{a priori} mean and variance of each frequency domain symbol are initialized by zero and $E_s$.
The frequency domain MMSE filter can then be written as
\begin{equation}
{{\bf{W}}}^{(i)} \buildrel \Delta \over = {{{\bf{\bar C}}}^{(i)}_z}{\bf{H}}^{\rm{H}}{\left( {{{\bf{H}}}{{{\bf{\bar C}}}^{(i)}_z}{\bf{H}}^{\rm{H}} + {N_0}{{\bf{I}}_{MN}}} \right)^{ - 1}},
\label{OTFS_freq_MMSE_full}
\end{equation}
where ${{{\bf{\bar C}}}^{(i)}_z} \buildrel \Delta \over = {\rm{diag}}\left( {{{{\bf{\bar v}}}^{(i)}_z}} \right)$ is the \emph{a priori} covariance matrix of $\bf z$.
Applying~\eqref{OTFS_freq_MMSE_full} to ${\bf q}$, we obtain the \emph{a posteriori} mean and error covariance matrix of $\bf z$ as
\begin{align}
{{{\bf{\hat z}}}}^{(i)} &= {{{\bf{\bar z}}}}^{(i)} + {{\bf{W}}}^{(i)}\left( {{{\bf{q}}} - {{\bf{H}}}{{{\bf{\bar z}}}}^{(i)}} \right), 
\label{OTFS_freq_MMSE_post_mean}\\
\vspace*{-0.5\baselineskip}
{{{\bf{\hat C}}}_z^{(i)}} &= {{{\bf{\bar C}}}^{(i)}_z} - {{\bf{W}}^{(i)}}{{\bf{H}}}{{{\bf{\bar C}}}^{(i)}_z}.
\label{OTFS_freq_MMSE_post_Cov}
\vspace*{-0.5\baselineskip}
\end{align}
For ease of implementation, we only consider the diagonal elements in the covariance matrix by exploiting the fact that the underlying variables are statistically uncorrelated, as it has been verified that imposing such a condition only causes marginal performance loss when $MN$ is sufficiently large~\cite{li2021cross}.
Let ${{\bf \hat v}^{(i)}_z}$ be the \emph{a posteriori} variance vector containing the diagonal entries of ${{{\bf{\hat C}}}^{(i)}_z}$.
After performing the FDE, the CDID passes ${{\bf{\hat z}}^{(i)}}$ and ${{\bf \hat v}^{(i)}_z}$ to the time domain and then to the DD domain for symbol-wise detection. Subsequently, the  DD domain \emph{a priori} mean vector 
and \emph{a priori} covariance matrix 
are respectively given by 
\begin{align}
{{{\bf{\bar x}}}}^{(i)} &= \left( {{{\bf{F}}_N} \otimes {{\bf{I}}_M}} \right){\bf{F}}_{MN}^{\rm{H}}{{{\bf{\hat z}}}}^{(i)},
\label{OTFS_DD_pri_mean} \\
\vspace*{-0.3\baselineskip}
{{{\bf{\bar C}}}^{(i)}_x} &= \left( {{{\bf{F}}_N} \otimes {{\bf{I}}_M}} \right){\bf{F}}_{MN}^{\rm{H}}{{{\bf{\hat C}}}^{(i)}_z}{{\bf{F}}_{MN}}\left( {{\bf{F}}_N^{\rm{H}} \otimes {{\bf{I}}_M}} \right),
\label{OTFS_DD_pri_cov}
\vspace*{-3mm}
\end{align}
whose diagonal entries can be rearranged in the vector form of ${{\bf \bar v}^{(i)}_x}$. 
Particularly, it can be shown that the variance ${{\bar v}^{(i)}_x}\left[ l \right]$ of the $l$-th DD domain symbol, for $0 \le l \le MN-1$, asymptotically converges to the mean of ${{\bf \hat v}^{(i)}_z}$~\cite{li2021cross}, i.e.,
\begin{align}
{{\bar v}^{(i)}_x}\left[ l \right] \simeq \frac{1}{{MN}}\sum\nolimits_{j = 0}^{MN - 1} {{{\hat v}^{(i)}_z}\left[ j \right]}  .
\label{OTFS_DD_pri_var}
\vspace*{-1\baselineskip}
\end{align}
Based on ${{{\bf{\bar x}}}}^{(i)}$ and ${{\bf {\bar v}}^{(i)}_x}$, we apply the symbol-wise detection.
Let ${\cal X}$ be the DD domain constellation set.  The \emph{a posteriori} probability (APP) of the $k$-th entry of ${\bf x}$ equals to the constellation point ${\cal X}_i$ can be calculated by
\vspace{-2mm}
{\small
\begin{equation}
\Pr \big( {{x}\left[ k \right] = {\cal X}_i|{{\bar x}}^{(i)}\left[ k \right]} \big) \propto \exp \Big( { - \frac{1}{{{{\bar v}^{(i)}_x}\left[ k \right]}}{{\left| {{\cal X}_i - {{\bar x}}^{(i)}\left[ k \right]} \right|}^2}} \Big).
\label{DD_APP}
\end{equation}}
\vspace{-3mm}

\noindent According to~\eqref{DD_APP}, the decision for ${x}\left[k\right]$ is selected as the constellation point maximizing the APP, which serves as the output for the current iteration.  We further calculate the \emph{a posteriori} mean ${{{\bf{\hat x}}^{(i)}}}$ and variance vector ${{\bf {\hat v}}^{(i)}_x}$ based on~\eqref{DD_APP} to enable the forthcoming iteration, which can be derived in a symbol-wise manner via
\vspace{-2mm}
{\small
\begin{align}
&{{\hat x}}^{(i)} \left[k\right]
\!=\! {\frac{1}{|{\cal X}|}}\!\!\sum\nolimits_{j = 0}^{|{\cal X}|-1}\Pr\left( x \left[k\right]={\cal X}_j|{{\bar x}}^{(i)}\left[k\right]\right){\cal X}_j,\label{mean_x_DD_post} ~~\text{and}
\\
&{{\hat v}^{(i)}_x}\left[ k \right] 
\!=\! \frac{1}{|{\cal X}|}\!\!\!\sum\nolimits_{j = 0}^{|{\cal X}| - 1}\!\! {\Pr }\! \left(\!\! {{x}\left[ k \right] \!=\! {\cal X}_j|{{\bar x}}^{(i)}\left[ k \right]} \right)\!\!{\left| {\cal X}_i \right|^2}\!\! -\!\! {\left| {{{\hat x}}^{(i)}\left[ k \right]} \right|^2}. \label{var_x_DD_post}
\end{align}}

\vspace{-2mm}
\noindent 
Finally, we transform ${{{\bf{\hat x}}}}^{(i)}$ and ${{\bf {\hat v}}^{(i)}_x}$ to the frequency domain to update ${\bar {\bf z}^{(i+1)}}$ and ${\bar {\bf v}}^{(i+1)}_{z}$ for the forthcoming iteration, i.e.,
\vspace{-1mm}
\begin{align}
{{{\bf{\bar z}}}}^{(i+1)} &= {\bf{F}}_{MN}\left( {{{\bf{F}}^{\rm H}_N} \otimes {{\bf{I}}_M}} \right){{{\bf{\hat x}}^{(i)}}}, ~~\text{and}
\label{OTFS_freq_pri_mean_next_iter}
\vspace*{-0.5\baselineskip} \\
{{{\bf{\bar C}}}_z}^{(i+1)}  &= {{\bf{F}}_{MN}}\left( {{\bf{F}}_N^{\rm{H}} \otimes {{\bf{I}}_M}} \right){{{\bf{\hat C}}}^{(i)}_x}\left( {{{\bf{F}}_N} \otimes {{\bf{I}}_M}} \right){\bf{F}}_{MN}^{\rm{H}}.
\label{OTFS_freq_pri_cov_next_iter}
\vspace*{-0.9\baselineskip}
\end{align}
Again, we rearrange the diagonal entries in ${{{\bf{\bar C}}}^{(i+1)}_z}$ into a vector ${{{\bf{\bar v}}}^{(i+1)}_z}$, which is passed to the FDE. Similar to~\eqref{OTFS_DD_pri_var}, we have
\begin{equation}
{{\bar v}^{(i+1)}_z}\left[ l \right] \simeq \frac{1}{{MN}}\sum\nolimits_{j = 0}^{MN - 1} {{{\hat v}^{(i)}_x}\left[ j \right]} .
\label{OTFS_freq_pri_var_next_iter}
\vspace*{-0.5\baselineskip}
\end{equation}

\noindent Based on~\eqref{OTFS_freq_pri_mean_next_iter} and~\eqref{OTFS_freq_pri_var_next_iter}, the next iteration can take place. 

\subsection{Type-II: CDID via Passing Extrinsic Information}
\begin{figure}
\centering
\includegraphics[trim=10bp 10bp 10bp 10bp,clip, width=0.4\textwidth ]{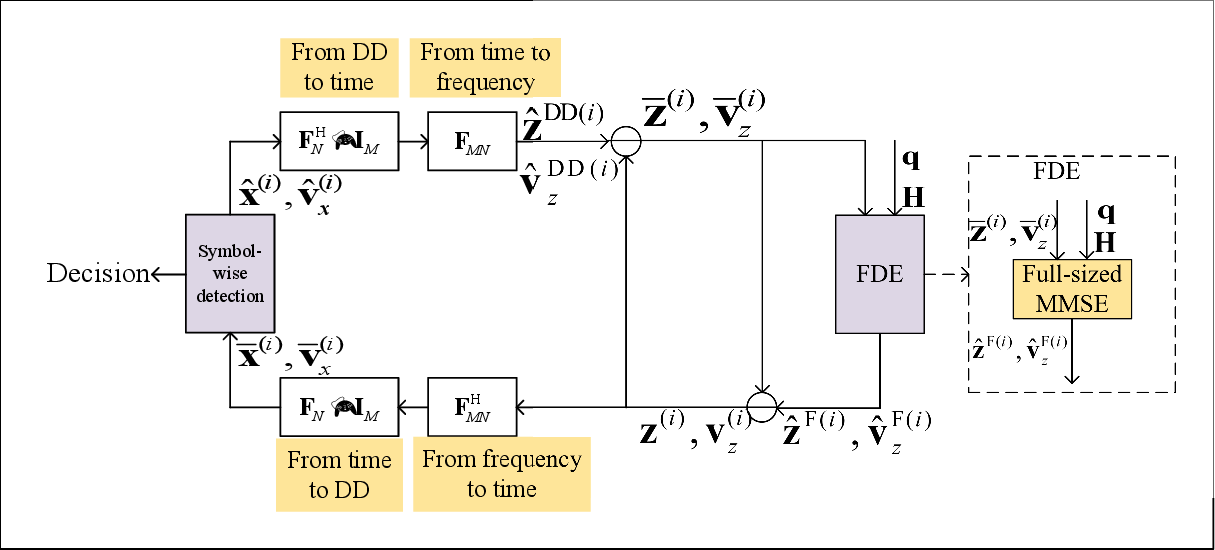}
\vspace{-3mm}
\caption{The block diagram for the Type-II CDID scheme.}
\label{cross_domain_typeIII}
\centering
\vspace{-5mm}
\end{figure}

The Type-II CDID differs from the Type-I CDID in terms of the message passing mechanism, where, instead of passing the \emph{a posteriori} information, the so-called ``extrinsic information'' is passed as shown in Fig.~\ref{cross_domain_typeIII}, which is derived based on both the \emph{a priori} and \emph{a posteriori} information.

To facilitate explanation, we slightly abuse the notations and refer to the \emph{a posteriori} mean and variance vectors of $\bf z$ derived from the FDE and DD domain detection in the $i$-th iteration by ${\bf{\hat z}}^{{\rm{F}}(i)},{\bf{\hat v}}_z^{{\rm{F}}(i)}$ and ${\bf{\hat z}}^{{\rm{DD}}(i)},{\bf{\hat v}}_z^{{\rm{DD}}(i)}$, respectively. Let ${\bf \tilde z}^{(i)}$ and ${\bf \tilde v}^{(i)}_{z}$ be the \emph{extrinsic} mean and variance vectors passed to the DD domain. 
Based on ${\bf \bar z} ^{(i)}$ and ${{\bf {\bar v}}^{(i)}_z}$, we have~\cite{Qinghua2011LMMSE}
\vspace{-2mm}
{\small
\begin{align}
{{\tilde v}^{(i)}_z}\left[ l \right] &= \frac{1}{{\frac{1}{{{{\hat v}_z^{{\rm F}{(i)}}}\left[ l \right]}} - \frac{1}{{{{\bar v}^{(i)}_z}\left[ l \right]}}}},~~\text{and}
\label{OTFS_ex_var_to_DD}
\vspace*{-0.5\baselineskip}\\
\tilde z^{(i)}\left[ l \right] &= {{\tilde v}^{(i)}_z}\left[ l \right]\left( {\frac{{\hat z^{{\rm F}(i)}\left[ l \right]}}{{{{\hat v}_z^{{\rm F}(i)}}\left[ l \right]}} - \frac{{\bar z^{(i)}\left[ l \right]}}{{{{\bar v}^{(i)}_z}\left[ l \right]}}} \right),
\label{OTFS_ex_mean_to_DD}
\vspace*{-0.5\baselineskip}
\end{align}}

\vspace{-2mm}
\noindent where ${{{\hat z}^{{\rm{F}}(i)}}\left[ l \right]}$ and ${{\hat v_z^{{\rm{F}}(i)}}\left[ l \right]}$ are the $l$-th entries of ${\bf{\hat z}}^{{\rm{F}}(i)}$ and ${\bf{\hat v}}_z^{{\rm{F}}(i)}$ obtained from the FDE in the current iteration.  
Similarly, the \emph{a priori} mean and variance vectors for the FDE are calculated following the same manner of~\eqref{OTFS_ex_var_to_DD} and~\eqref{OTFS_ex_mean_to_DD} as
\vspace{-1mm}
\begin{align}
{{\bar v}^{(i)}_z}\left[ l \right] &= \frac{1}{{\frac{1}{{\hat v_z^{{\rm{DD}}(i)}\left[ l \right]}} - \frac{1}{{{{\tilde v}^{(i)}_z}\left[ l \right]}}}},~~\text{and}
\vspace*{-0.5\baselineskip}
\label{OTFS_ex_var_to_freq}\\
\bar z^{(i)}\left[ l \right] &= {{\bar v}^{(i)}_z}\left[ l \right]\left( {\frac{{{{\hat z}^{{\rm{DD}}(i)}}\left[ l \right]}}{{\hat v_z^{{\rm{DD}}(i)}\left[ l \right]}} - \frac{{\tilde z^{(i)}\left[ l \right]}}{{{{\tilde v}^{(i)}_z}\left[ l \right]}}} \right),
\label{OTFS_ex_mean_to_freq}
\vspace*{-0.5\baselineskip}
\end{align}
where ${{{\hat z}^{{\rm{DD}}(i)}}\left[ l \right]}$ and ${{\hat v_z^{{\rm{DD}}(i)}}\left[ l \right]}$ are the $l$-th entries of ${\bf{\hat z}}^{{\rm{DD}}(i)}$ and ${\bf{\hat v}}_z^{{\rm{DD}}(i)}$. 
The motivation for passing the extrinsic information in the Type-II CDID algorithm is to ensure that the outputs from different modules are not directly fed back to themselves during successive iterations, which is known to improve the robustness of general iterative algorithms. 

\section{Performance Analysis}
In this section, we pursue a performance analysis for the proposed CDIDs. Due to the page limitation, we only present the main conclusions in the following without explicitly highlighting the derivations. The detailed derivations of related conclusions can be found in the journal version of this paper.

To better understand the proposed algorithm, it is important to study the foundational intuition of the algorithm first. Recall that the proposed CDIDs apply different forms of FDE while employing symbol-by-symbol detection in the DD domain in each iteration. Evidently, these algorithms are effective only when the DD domain's effective channel after FDE, can be closely approximated by ${\bf \bar x}^{\left(i\right)}={\bf x}+{\bm \varepsilon }^{\left(i\right)}$, where ${\bm \varepsilon }^{\left(i\right)}$ is a vector of effective noise in the $i$-th iteration that can be roughly treated as a vector of zero-mean white Gaussian variables with known variance ${\bar v}_x^{\left(i\right)}$. It can be shown that the above condition holds generally when the adopted FDE is unbiased, as discussed in the journal version of this paper.{\footnote{Note that this is only a sufficient condition but not a necessary condition as explained in our journal paper.}} Therefore, we first evaluate the estimation bias in the following subsection.

\subsection{Bias Evolution}
\begin{figure}
\centering
\includegraphics[scale=0.4]{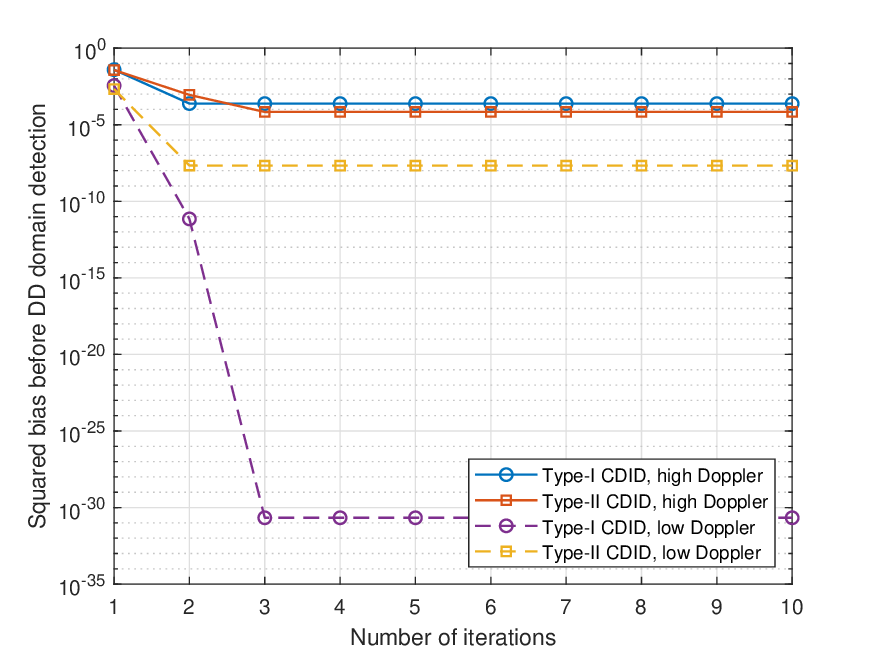}
\vspace{-3mm}
\caption{Evaluation of estimation bias before DD domain detection.}
\label{bias:2}
\vspace{-4mm}
\end{figure}

A mathematical derivation for the average bias of each iteration may be difficult due to the non-linear nature of the DD domain detection{\footnote{Theoretically, the effect of the DD domain detection can be determined by taking the two-dimensional integral over the complex domain of the Euclidean distances between the estimation outputs and the constellation points.}} and the coupling with the estimated variance. Therefore, we propose to evaluate the DD domain estimation bias in the two domains recursively using Monte-Carlo simulations as follows. We consider a given frequency domain channel matrix ${\bf H}$ and an energy-normalized constellation set $\cal X$. In each Monte-Carlo trial, we generate the DD domain symbol vector $\bf x$, whose entry is obtained by equal-probably taking values from $\cal X$. Then, the frequency domain symbol vector $\bf z$ is obtained by ${\bf z}={\bf U}^{\rm H}{\bf x}$. For each 
$\bf x$, we generate 
${\mathbb E}\left\{{\bf \bar x}^{\left(i\right)}\right\}$ by executing the proposed algorithms multiple times.
We consider $M=32$, $N=16$, $P=4$ and the transmit ${\rm SNR}\triangleq \frac{E_s}{N_0}=15$ dB. The channel fading coefficients and delay indices are set as $[0.5, 0.5, 0.5, 0.5]$, and $[0, 1, 3, 1]$, respectively. Moreover, the Doppler indices are set as $[0.95, 4.9, 2.2, -1.5]$ for high-Doppler cases and $[0.2, 0.15, -0.18, -0.08]$ for low-Doppler cases.
From the illustrated squared estimation bias, which is defined by ${{\rm bias}\left( {{{{\bf{\bar x}}}^{\left( i \right)}}\!,{\bf{x}}} \right)}\triangleq{{\mathbb E}\left\{ {{{{\bf{\bar x}}}^{\left( i \right)}}} \right\} \!\!-\! {\bf{x}}}$ between ${{{\bf{\bar x}}}^{\left( i \right)}}$ and $\bf x$, in 
Fig.~\ref{bias:2}, we observe that both Type-I and Type-II CDIDs suffer from only negligible estimation bias in the DD domain.
Intuitively, this is because the MMSE estimators are known to be unbiased when the \textit{a priori} mean and covariance matrix are accurate.

\vspace{-1mm}
\subsection{State Evolution}
In this subsection, we study the estimation variance of the proposed CDIDs. This study is meaningful particularly when the underlying estimation bias is negligible, such that the average MSEs are dominated by the estimation variance.
To this end, we define the frequency domain error state by the trace of the covariance matrix of ${\bf \bar z}^{\left(i\right)}$, i.e.,
\vspace{-1mm}
{\small
\begin{align}
\eta _z^{\left( i \right)} \triangleq \frac{1}{{MN}}{\rm Tr}\left({\bf \bar C}_z^{\left( i \right)}\right)=\frac{1}{{MN}}\sum\nolimits_{l = 0}^{MN - 1} {\bar v_z^{\left( i \right)}\left[ l \right]} .
\label{Freq_state}
\end{align}}

\noindent Similarly, the DD domain error state is defined by 
\vspace{-1mm}
{\small
\begin{align}
\eta _x^{\left( i \right)} \buildrel \Delta \over  = \frac{1}{{MN}}{\rm Tr}\left({\bf \bar C}_x^{\left( i \right)}\right)=  \frac{1}{{MN}}\sum\nolimits_{j = 0}^{MN - 1} {\bar v_x^{\left( i \right)}\left[ j \right]}  .
\label{DD_state}
\end{align}}

\noindent Let us consider the $i$-th iteration, and define ${{\bf{P}} \triangleq {{\bf{H}}}{\bf{H}}^{\rm{H}}}$, whose eigenvalue decomposition is ${{\bf{P}}} = {\bf{{\tilde U}\Lambda }}{{\bf{\tilde U}}^{\rm{H}}}$, where $\bf \tilde U$ is a unitary matrix and $\bf \Lambda$ is a diagonal matrix containing the eigenvalues of ${{\bf{P}}}$, i.e., ${\bf{\Lambda }} = {\rm{diag}}\left\{ {\left[ {{\lambda _0},{\lambda _1},...,{\lambda _{MN - 1}}} \right]} \right\}$. 
Furthermore, we define an arbitrary function $g\left( {\cdot} \right)$ to characterize the relation between $\hat v_x^{\left( i \right)}\left[ j \right]$ and $\bar v_x^{\left( i \right)}\left[ j \right]$, i.e., $\hat v_x^{\left( i \right)}\left[ j \right] = g\left( {\bar v_x^{\left( i \right)}\left[ j \right]} \right)$, which is assumed to be non-increasing and non-negative. Here, we contend that the aforementioned assumption is justified by the fact that a competent detector can typically diminish the uncertainty associated with non-Gaussian constellations.
In fact, the approximations were also adopted in~\cite{li2021cross}, which were shown to be sufficient for predicting the overall error performance of CDIDs under various channel conditions.
With the above, we have the following propositions. 

\textbf{Proposition 1} (\emph{State Evolution for the Type-I CDID}):
For Type-I CDID, the DD domain error state converges to
\vspace{-2mm}
{\small
\begin{equation}
\vspace{-1.5mm}
\eta _x^{\left( i \right)} 
\simeq\eta _z^{\left( i \right)} - \frac{{{{| {\eta _z^{\left( i \right)}} |}^2}}}{{MN}}\sum\nolimits_{l = 0}^{MN - 1} {\frac{{{\lambda _l}}}{{\eta _z^{\left( i \right)}{\lambda _l} + {N_0}}}} .
\label{SE_1}
\end{equation}}
Moreover, the frequency domain error state converges to
{\small
\begin{equation}
\eta _z^{\left( {i + 1} \right)} \simeq
  \frac{1}{{MN}}\sum\nolimits_{l = 0}^{MN - 1} {\hat v_x^{\left( i \right)}\left[ l \right]} 
= {\mathbb E}\left\{ {g\left( {\eta_x^{\left( i \right)}} \right)} \right\}.
\label{SE_der3}
\end{equation}}

\textbf{Proposition 2} (\emph{State Evolution for the Type-II CDID}):
For Type-II CDID, the DD domain and frequency domain error states converge to
{\small
\begin{align}
\eta _x^{\left( i \right)} 
&\simeq
{\bigg ({\frac{1}{\eta _z^{\left( i \right)} -  \frac{{{{| {\eta _z^{( i )}} |}^2}}}{{MN}}\sum\nolimits_{l = 0}^{MN - 1}\!\!\! {\frac{{{\lambda _l}}}{{\eta _z^{\left( i \right)}{\lambda _l} + {N_0}}}}} - \frac{1}{{\eta _z^{(i)} }}}\bigg )}^{-1},
\label{SE_der7}\\
\eta _z^{\left( {i + 1} \right)} 
&\simeq 
{\bigg (\frac{1}{{\mathbb E}\big\{ {g\big( {\eta _x^{( i )}} \big)} \big\}} - \frac{1}{\eta _x^{( i )}}\bigg )}^{-1}.
\label{SE_der8}
\end{align}
}

From the above propositions, we notice that the eigenvalues of $\bf P$ significantly influence the parameter $\eta _x^{\left( i \right)}$. To obtain further insights, we apply Jensen's inequality and exploit the fact that $ \frac{1}{{MN}}\sum\nolimits_{l = 0}^{MN - 1} {{\lambda _l}}  \approx \sum\nolimits_{i = 1}^P {{{\left| {{h_i}} \right|}^2}} $, when different resolvable paths have different delay indices~\cite{Ruoxi2022achievable}.\footnote{It should be noted that this assumption is reasonable in practice as the resolvable paths usually come from geographically distributed reflectors and therefore have different path delays.} As such, the following corollary holds naturally.

\textbf{Corollary 1} \emph{(Lower bounds for DD Domain Error State)}:
For the Type-I CDID, we have
\begin{align}
\eta _x^{\left( i \right)} 
&\ge \frac{{\eta _z^{\left( i \right)}{N_0}}}{{\eta _z^{\left( i \right)}\sum\nolimits_{i = 1}^P {{{\left| {{h_i}} \right|}^2}}  + {N_0}}}
=\frac{N_0}{\sum\nolimits_{i = 1}^P {{\left| {{h_i}} \right|}^2}  + \frac{N_0}{\eta _z^{\left( i \right)}}}.
\label{lower_bound_SE_I}
\end{align}

\noindent Similarly, for the Type-II CDID, we have 
\begin{align}
\eta _x^{\left( i \right)} \ge 
\frac{N_0}{\sum\nolimits_{i = 1}^P {{\left| {{h_i}} \right|}^2} }.
\label{lower_bound_SE_2}
\end{align}

\subsection{Analysis of the Error Performance in Convergence}

\begin{figure*}
\begin{minipage}{0.32\linewidth}
  \vspace{-1mm}
  \centerline{\includegraphics[width=\textwidth]{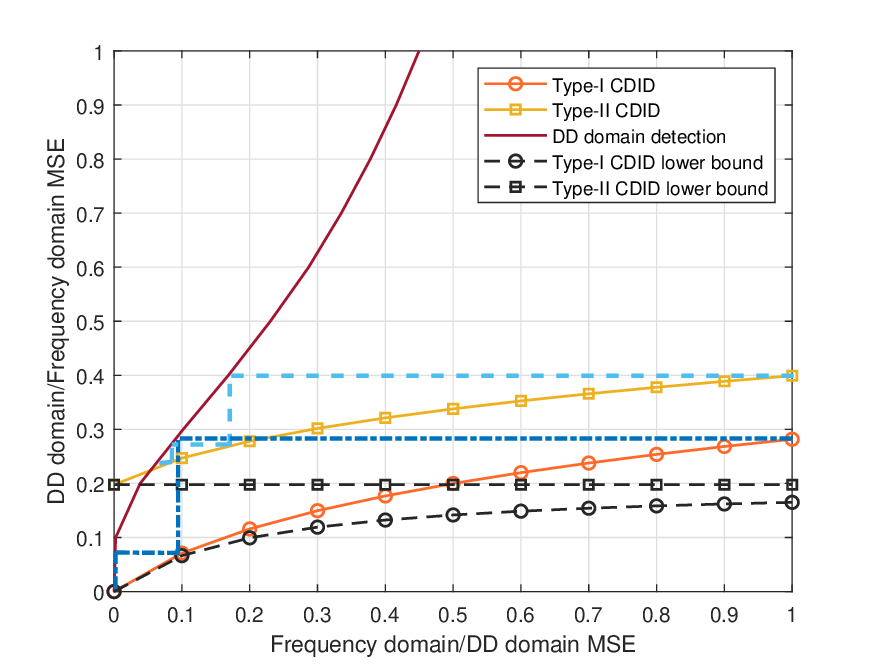}}
  \vspace{-3mm}
  \caption{Heuristic state evolution and convergence trajectory with high Doppler shift at transmit ${\rm SNR}=15$ dB.
  }
  \label{CCDF}
\end{minipage}
\begin{minipage}{0.32\linewidth}
  \centerline{\includegraphics[width=\textwidth]{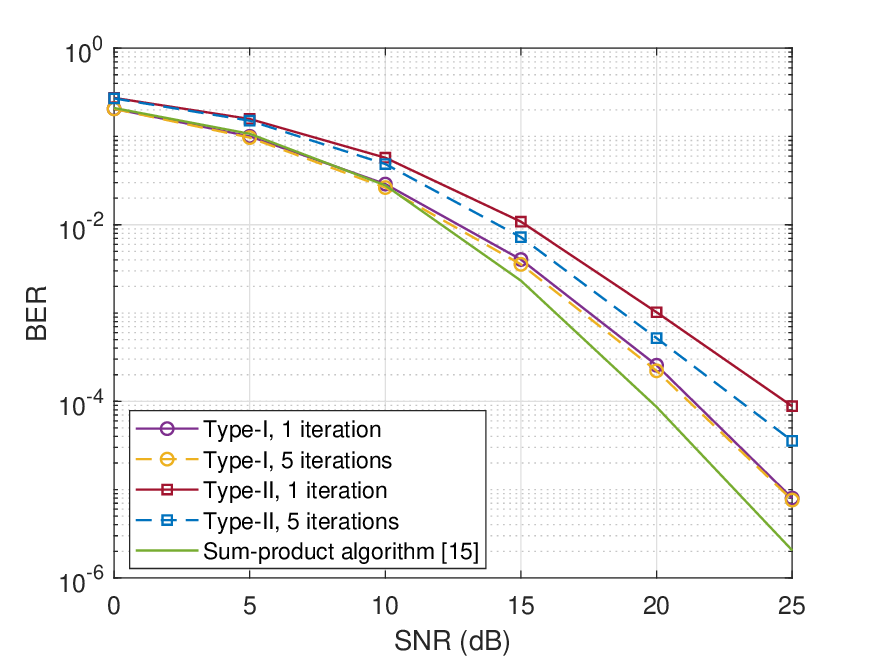}}
  \vspace{-2mm}
  \caption{Performance comparison between Type I and Type II CDID schemes with different iterations in negligible Doppler shift scenarios.}
  \label{fig:1}
\end{minipage}
\begin{minipage}{0.32\linewidth}
  \centerline{\includegraphics[width=\textwidth]{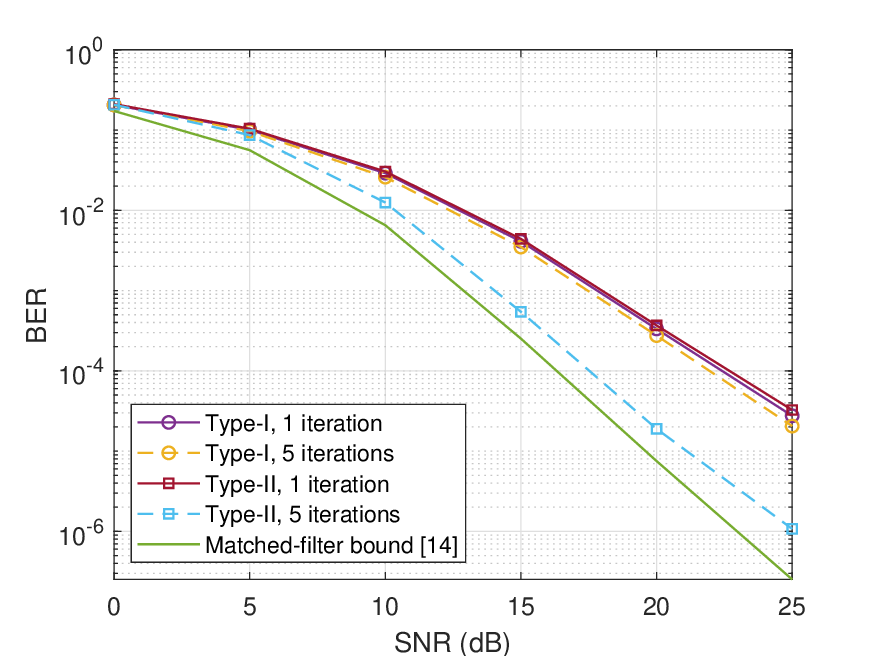}}
  \vspace{-2mm}
  \caption{Performance comparison between Type-I and Type-II CDID schemes with different iterations in high Doppler shift scenarios.}
  \label{fig:2}
\end{minipage}
\vspace{-3.5mm}
\end{figure*}

When the algorithm converges, fixed points of error states can be achieved. The proof of the existence of fixed points is important, but we omit it here due to space limitations. However, this will be verified by our numerical results later.
By considering $\eta_x^{\star}$ and $\eta_z^{\star}$ as the DD domain and frequency domain error states in convergence. 
For the Type-I CDID in convergence, we have
\begin{align}
\eta _x^{\star} \ge \frac{{\eta _z^{\star}{N_0}}}{{\eta _z^{\star}\sum\nolimits_{l = 1}^P {{{\left| {{h_i}} \right|}^2}}  + {N_0}}}.
\label{SE_FP_der1}
\end{align}
By substituting~\eqref{SE_der3} into~\eqref{SE_FP_der1}, we see that to satisfy~\eqref{SE_FP_der1}, one of the following conditions must hold:  
\begin{subequations}
\begin{align}
&{\mathbb E}\{g\left( {\eta _x^{\star}} \right)\} \!\le\! \frac{{{N_0}\eta _x^{\star}}}{{{N_0} \!- \!\eta _x^{\star}\sum\nolimits_{l = 1}^P {{{\left| {{h_i}} \right|}^2}} }}, 
\quad{\text{for}~{\eta _x^{\star}} \!<\! \frac{N_0}{{\sum\nolimits_{l = 1}^P {{{\left| {{h_i}} \right|}^2}} }}}, \label{a}\\
&{\mathbb E}\{g\left( {\eta _x^{\star}} \right)\} \!\ge\! \frac{{{N_0}\eta _x^{\star}}}{{{N_0} \!-\! \eta _x^{\star}\sum\nolimits_{l = 1}^P {{{\left| {{h_i}} \right|}^2}} }},
\quad{\text{for}~{\eta _x^{\star}} \!>\! \frac{N_0}{{\sum\nolimits_{l = 1}^P {{{\left| {{h_i}} \right|}^2}} }}}. \label{b}
\end{align}
\end{subequations}

\noindent Note that both conditions above are possible and therefore the convergence of Type-I CDID does not depend on the estimation quality of the FDE since convergence will be achieved regardless of the ${\eta _x^{\star}}$ value as suggested in~\eqref{a} and~\eqref{b}. However, its ultimate error performance is strongly dependent on the accuracy of the FDE. Particularly, the algorithm may be likely to converge to an incorrect decision if the FDE estimate is not sufficiently accurate, as it keeps passing the \textit{a posteriori} information. On the other hand, if the FDE estimate is very accurate, the Type-I CDID is likely to provide good error performance. This may happen when the frequency domain channel matrix is fully diagonal, i.e., the channel has no Doppler shift, and the LMMSE estimator is near-optimal for maximizing the likelihood function.

For the Type-II CDID scheme, based on~\eqref{SE_der7},~\eqref{SE_der8}, and Jensen's inequality, after some manipulations, we arrive at
{\small
\begin{align}
    {{\mathbb E}\left\{ {g\left( {\eta _x^{{\star}}} \right)} \right\}} \bigg(\frac{N_0}{{\sum\nolimits_{l = 1}^P {{{\left| {{h_i}} \right|}^2}} }} - {\eta_x^{\star}}\bigg) \ge {\eta_x^{\star}} \bigg(\frac{N_0}{{\sum\nolimits_{l = 1}^P {{{\left| {{h_i}} \right|}^2}} }} - {\eta_x^{\star}}\bigg).
    \label{TYPE2_ineq}
\end{align}}

\noindent Therefore, for the Type-II CDID in convergence, the following constraints must be satisfied:
{\small
\begin{subequations}
\begin{align}
&{\mathbb E}\{g\left( {\eta _x^{\star}} \right)\} \ge \eta _x^{\star}, 
\quad{\text{for}~{\eta _x^{\star}} < \frac{N_0}{{\sum\nolimits_{l = 1}^P {{{\left| {{h_i}} \right|}^2}} }}} ,\label{a_2}\\
&{\mathbb E}\{g\left( {\eta _x^{\star}} \right)\} \le \eta _x^{\star},
\quad{\text{for}~{\eta _x^{\star}} > \frac{N_0}{{\sum\nolimits_{l = 1}^P {{{\left| {{h_i}} \right|}^2}} }}} .\label{b_2}
\end{align}
\end{subequations}}

\noindent 
Note that~\eqref{a_2} is not achievable based on~\eqref{lower_bound_SE_2}. Therefore, in convergence, the frequency domain error state of the Type-II CDID is lower than $N_0/{{\sum\nolimits_{l = 1}^P {{{\left| {{h_i}} \right|}^2}} }}$. Notice that $N_0/{{\sum\nolimits_{l = 1}^P {{{\left| {{h_i}} \right|}^2}} }}$ is the error variance corresponding to the matched filter bound~\cite{Mazo1991MFB}, which indicates the optimal FDE performance. 
Therefore, the analytical results suggest that the Type-II CDID can achieve a promising detection performance in convergence. This is not unexpected, because the extrinsic information is known for improving the robustness against error propagation, thereby, leading to a better error performance~\cite{Qinghua2011LMMSE}.

\vspace{-1mm}
\section{Numerical Results}
\vspace{-1mm}
In this section, we evaluate the bit error rate (BER) performance and validate our analytical findings via simulations.
We consider the rectangular pulse-shaped OTFS transmission with $M=32$, $N=16$, and $L_{\rm CP}=l_{\max}=3$, where $l_{\max}$ is the maximum delay index. We consider the QPSK constellation, $P=4$ resolvable paths, and the channel delay and Doppler indices can admit fractional values.
As benchmarks for the BER, we also plot the matched filter bound~\cite{Mazo1991MFB} and the sum-product algorithm~\cite{li2021hybrid} performance, where the sum-product algorithm is implemented by assuming integer delay and Doppler indices in order to maintain a feasible complexity.

In Fig.~\ref{CCDF}, we present the state evolution and the heuristic converge trajectories at SNR $=15$ dB for both Type-I and Type-II CDID algorithms implemented in a high Doppler shift channel for bias evolution.
The labels of the $x$ and $y$ axes denote the input-output MSE pairs in either the frequency or DD domain.
From the figure, we observe that both CDIDs can converge at low MSEs, and the derived lower bounds match well with the average MSE for both the DD and frequency domains. 
Specifically, the error states of Type-I CDID can approach zero while the error states of Type-II CDID are lower-bounded by the matched filter bound. This aligns with our previous discussion on Corollary 1. 

Figure~\ref{fig:1} depicts the BER performance of the considered CDIDs QPSK constellations, where the maximum Doppler index is $k_{\max}=0.2$. 
From the figure, 
we observe that Type-I CDID outperforms Type-II CDID for both one and five iterations.  
This is because, in the presence of negligible Doppler shifts, the frequency domain channel is nearly diagonal, and thus the performance of FDE approaches the optimal. Therefore, passing the \textit{a posteriori} information can result in sufficiently good error performance, while passing the extrinsic information degrades the performance due to the poor quality of the extrinsic information due to the negligible energy of the no-diagonal elements.
Furthermore, we notice that the performance of CDID algorithms approaches the near-optimal sum-product algorithm in the low SNR regime with reduced complexity, even in the presence of fractional delay and Doppler shifts. 

Figure~\ref{fig:2} illustrates the BER performance for the considered CDIDs with a maximum Doppler index $k_{\max}=5$. We notice that the performance improvement for Type-I CDID due to iterations is marginal. In contrast, the Type-II CDID exhibits a significant performance improvement with additional iterations and also yields a promising performance approaching that of the matched filter bound, especially in the low SNR regime.
This is attributed to the robustness of the extrinsic information against the error propagation.

\vspace{-1mm}
\section{conclusion}
\vspace{-1mm}
We investigated the performance of two types of CDID algorithms with different information passing mechanisms in the context of OTFS modulation. To analyze the system performance, both bias and error state evolutions were derived. Moreover, the error state in convergence was discussed to demonstrate the achievable error performance. 
Our simulation results corroborated our theoretical discussions and demonstrated a promising error performance.

\bibliographystyle{IEEEtran}
\bibliography{OTFS_references}
\end{document}